# Lattice and polarizability mediated spin activity in EuTiO$_3$

K. Caslin[1,2], R. K. Kremer[1], Z. Guguchia[3], H. Keller[3], J. Köhler[1], A. Bussmann-Holder[1]

[1]Max Planck Institute for Solid State Research, Heisenbergstr.1, D-70569 Stuttgart, Germany
[2]Brock University, 500 Glenridge Ave., St. Catharines L2S-3A1, Ontario, Canada
[3]Physik-Institut der Universität Zürich, Winterthurerstr. 190, CH-8057 Zürich, Switzerland

EuTiO$_3$ is shown to exhibit novel strong spin-charge-lattice coupling deep in the paramagnetic phase. Its existence is evidenced by an, until now, unknown response of the paramagnetic susceptibility at temperatures exceeding the structural phase transition temperature $T_S$=282K. The "extra" features in the susceptibility follow the rotational soft zone boundary mode temperature dependence above and below $T_S$. The theoretical modeling consistently reproduces this behavior and provides reasoning for the stabilization of the soft optic mode other than quantum fluctuations.

Pacs-Index 75.30.Kz, 75.95.+t, 63.70.+h

Multiferroics have been in the focus of intense research since long with a revival during the last decade. The complex properties of these materials allows for the ability to tune different functionalities by various tools and thus to open a vast potential for applications. Perovskites oxides are typically at the forefront in this research area since they combine multiple ground states including ferro- and antiferromagnetism, ferroelectricity, and superconductivity. While ferroelectric properties are known to be related to the transition metal B d$^0$ state in ABO$_3$, magnetic properties are generally achieved in systems where the B-site has a d$^n$ (n>0) configuration. Obviously, this seems to exclude the simultaneous occurrence of magnetism and ferroelectricity. An interesting way out of this dilemma is to retain the d$^0$ configuration at the B site intact and to place a magnetic ion at the A site. This has been realized in EuTiO$_3$ (ETO) which was first synthesized more than half a century ago [1] and later considered in the search for new ferroelectrics [2]. In view of the absence of polar order, this compound has not been investigated any deeper but attracted novel interest only recently, when it was shown, that the dielectric constant of this material exhibits an unusual drop at the onset of antiferromagnetic order at $T_N$=5.7K [3] which can be reversed by a magnetic field. Above $T_N$ the dielectric constant decreases with increasing temperature which has been attributed to the softening of a long wave length transverse optic mode, reminiscent of a polar instability [4, 5]. The obvious analogy of this mode softening with similar behavior in SrTiO$_3$ (STO) [6] raised speculations that quantum fluctuations are responsible for the incomplete mode softening thus classifying ETO as a quantum paraelectric [7].
More recently, the analogies between ETO and STO have been expanded by the common structural phase transition related to the oxygen octahedral rotation which is caused by the condensation of a transverse acoustic zone boundary mode [8, 9]. This newly discovered phase transition together with the low temperature transition in ETO has been intensively investigated theoretically and experimentally [8, 9, 11 -19], and in particular commonalities between both have been discussed [8 – 10, 20 – 22]. Theoretically it has been suggested that the same modeling for both systems, ETO and STO, are at work, where for the former an additional spin-lattice coupling term was necessary in the model to account for the antiferromagnetic phase transition and its ramifications on the dielectric constant [8 – 10]. Also, biquadratic and higher order spin-lattice interaction terms have been included in either describing the lattice or the spin responses [3, 23, 24]. The focus of this latter work concentrates on the low temperature anomaly of the dielectric constant and the magnetic field dependence of the magnetic susceptibility.

By carrying out electron paramagnetic resonance (EPR) and muon spin rotation (µSR) measurements on ETO at high temperatures, we have demonstrated that active correlated spin dynamics must be considerably present at high temperatures exceeding the structural phase transition temperature $T_S$ [16, 18, 25]. These experiments have evidenced that the paramagnetic µSR signal as well as the inverse EPR line width are strongly coupled to the zone boundary acoustic mode dynamics. A further demonstration of this coupling was achieved by establishing a strong magnetic field dependence of $T_S$ [18].

Here we start from theoretical modeling of spin-lattice coupled systems [8] which we propose is relevant not only for ETO but also for many other compounds where lattice, spin and electronic degrees of freedom are interrelated and lead to concerted actions. For the lattice we use the polarizability model Hamiltonian [26 – 28], where charge degrees of freedom are simulated through electronic shells surrounding the polarizable $TiO_3$ cluster with mass $m_1$. The Eu ions with mass $m_2$ are coupled harmonically through the coupling constant $f$ to the shells of $m_1$. The core-shell coupling is nonlinear consisting of an attractive harmonic part $g_2$ and a repulsive stabilizing fourth order term $g_4$. The overall lattice stability is assured by the second nearest neighbor core-core coupling $f'$ between the polarizable units:

$$H = \sum_i \frac{1}{2}[m_i \dot{u}_{in}^2 + f'(u_{1n+1} - u_{1n})^2 + g_2(v_{1n} - u_{1n})^2 + \frac{1}{2} g_4 (v_{1n} - u_{1n})^4 + \\ f(u_{2n} - v_{1n})^2 + f(u_{2n+1} - v_{1n})^2] \quad (1)$$

with $u_{in}$, $v_{1n}$ being the displacement coordinates of ion $i$ and shell 1 in the n'th unit cell. This Hamiltonian has been shown to be very successful in describing ferroelectric perovskite oxides where the self-consistent phonon approximation (SPA) as well as nonlinear solutions have been studied and applied [26 – 31]. The spin Hamiltonian considers two exchange interactions: the antiferromagnetic $J_1$ between nearest neighbor spins, and the ferromagnetic $J_2$ between second nearest neighbors (Fig. 1):

$$H = z_1 \sum_{\langle i,j \rangle} J_1 S_i S_j + z_2 \sum_{[i,j]} J_2 S_i S_j - \sum_i h S_i, \quad (2)$$

where $z_1$, $z_2$ are the numbers of nearest and next-nearest neighbors, $S$ the Eu spins ($S=7/2$) and $h$ is an external field.

While we have used in our previous work an effective linear coupling between spin and lattice degrees of freedom [8], we consider here a quadratic coupling of the shell displacements to the spins [23, 24]:

$$H = \alpha \sum_{n,i,j} v_{1n}^2 S_i S_j \quad (3)$$

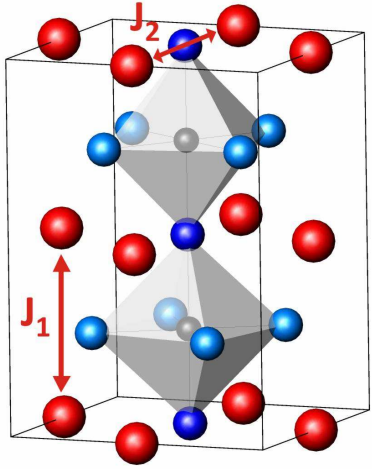

**Fig. 1** (color online) The tetragonal structure of EuTiO$_3$ where the relevant spin interactions $J_1$, $J_2$ are inserted as red arrows. The red balls are Eu, the blue ones oxygen, and the grey ones Ti ions, respectively. The tetragonal distortion has been exaggerated in order to make this clearly visible.

The coupling α between shell and spins stems from the intermediate oxygen ion and modifies the second nearest neighbor exchange in ETO. Note, that in other magnetic systems the nearest neighbor exchange interaction is affected by α if the oxygen lies between nearest neighbor magnetic ions. The consequences of this coupling are nevertheless similar. Since we are interested here in the interplay between spin, charge and lattice in ETO, in the following $J_2$ is renormalized through α leading to the modified spin Hamiltonian:

$$H = z_1 \sum_{\langle i,j \rangle} J_1 S_i S_j + z_2 \sum_{[i,j]} (J_2 + \alpha \langle w \rangle_T^2) S_i S_j - \sum_i h S_i . \qquad (4)$$

Importantly, it has to be noted that the spin-charge coupling enhances the ferromagnetic term and introduces a strong temperature dependence through $<w_{1n}^2>_T$ which is the thermal average of the relative core shell displacement coordinate $w = v - u$ squared corresponding to a polarizability coordinate. Within the SPA, this quantity is explicitly given by:

$$g_T = g_2 + 3g_4 \langle w \rangle_T^2 = \sum_{i,q} \frac{\hbar}{m\omega_i(q)} W^2(q) \coth \frac{\hbar \omega_i(q)}{2k_B T} \qquad (5)$$

where the sum is over all momentum $q$ dependent phonon branches $i$ and $W(q)$ is the eigenvector of the polarizability coordinate. In ferroelectric perovskites $g_T = 0$ marks the transition to the polar state and defines implicitly the transition temperature T$_C$ [26 – 28]. Within the sum in eq. 5 the largest contributions to $g_T$ stem from soft modes due to their pronounced temperature dependence, and it has been shown previously [26] that $\langle w \rangle_T^2$ follows essentially the temperature dependence of this mode.

The spin induced lattice alterations add the additional term Eq. (3) to Eq. (1) and modify substantially the dynamics of the coupled system which has been discussed before [8, 9, 18]. Here we consider in more detail the consequences for the spin system. An interesting implication, which is novel and has to our knowledge, never been discussed before, is a lattice induced enhancement of the magnetic susceptibility which should trace the soft mode temperature dependence. This is evident in the equation for the magnetic susceptibility which in the high temperature limit deep in the paramagnetic phase is given by:

$$\chi = N_A \mu_B^2 g^2 S(S+1)(1+\alpha \langle w \rangle_T^2)/(3k_B T) \quad (6)$$

where $N_A$ is the Avogadro number, $g$ the Landé factor, $\mu_B$ the Bohr magneton. Depending on the strength of the coupling constant α, a significant increase in the magnetic susceptibility as compared to the uncoupled case is expected. This is, however, not the major effect to demonstrate the existence of spin lattice coupling, more important is the *additional temperature dependence* of this enhancement which should reflect the soft mode temperature dependence. Since the transverse optic long wave length mode softening is irrelevant for this spin-lattice coupling, the rotational oxygen octahedral mode instability has to be considered in which the Eu – Eu bridging oxygen ions are involved. Its temperature dependence has been calculated within the same approach as used here and almost perfectly displacive dynamics have been observed [9, 25]. The contribution of this mode to the magnetic susceptibility should be minimal at the phase transition temperature and increase systematically below and above $T_S$. In order to demonstrate the expected extra contribution, the squared soft zone boundary acoustic mode frequency is shown as a function of (T-$T_S$) in Fig. 2 where renormalization effects stemming from the spin-lattice coupling [Eq. (3)] have not been included.

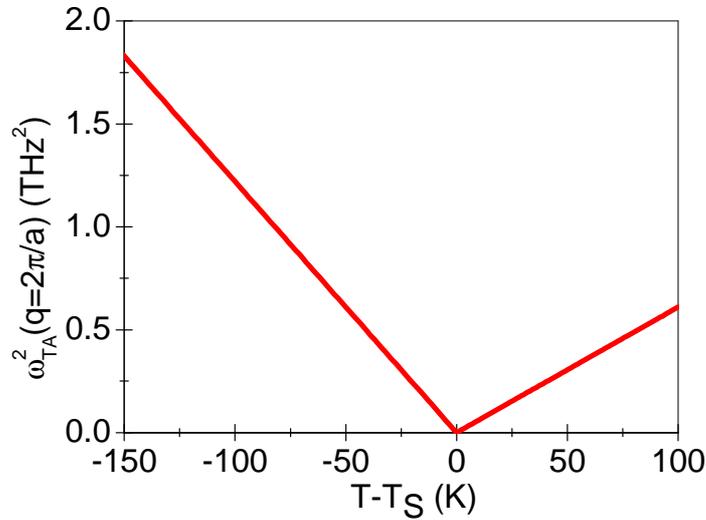

**Fig. 2** The squared soft zone boundary acoustic mode as a function of (T-$T_S$) [24].

The magnetic susceptibility of several polycrystalline samples of ETO has been measured with a *Quantum Design* magnetic properties measurement system (MPMS) SQUID magnetometer over a large temperature region and is shown in Fig. 3. The overall temperature dependence of χ can well be described with a Curie-Weiss law: $\chi = C/(T - \Theta_W)$ where $C$=7.7 (1) K is the Curie constant and $\Theta_W$=3.5.(1) K the Cuire-Weiss temperature.

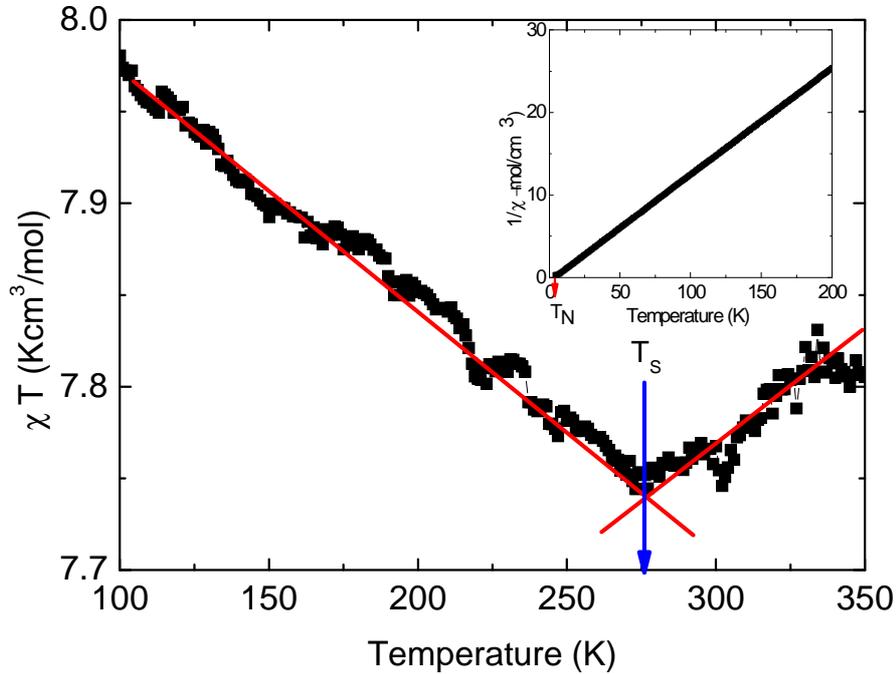

**Fig. 3** Temperature dependence of χT in a temperature interval between 100 and 350K. The intercept between the high and low temperature linear regime marks the structural instability at $T_S$ as indicated by the arrow. The red lines are a guide to the eye. The inset shows the temperature dependence of the inverse magnetic susceptibility $1/\chi$ over a large temperature region.

Obviously, no distinct anomalies are seen on this scale. However, as is evident from Eq. (3) anomalies should be visible if $3k_BT\chi = N_A\mu_B^2 g^2 S(S+1)(1+\alpha\langle w\rangle_T^2)$ is plotted since the first term on the right hand side is temperature independent and only from the last term an additional temperature dependence can be expected. This has been shown in Fig. 3 which clearly demonstrates that the magnetic susceptibility traces the structural instability and exhibits a temperature dependence following the one of the transverse acoustic zone boundary mode (Fig. 2). $T_S=278(5)$K extracted from the intercept of the linear in T relations above and below $T_S$, is close to the formerly reported value of $T_S=282$K [8]. This small deviation is most likely attributable to tiny sample preparation differences. The results, however, are a clear indication, that spin-lattice coupling is very strong in ETO and that magnetic susceptibility data are a suitable tool to detect these features. The results also show that it is possible to locate very precisely a structural phase transition temperature without the involvement of long range magnetic order. It is, however, anticipated that short range fluctuating ferromagnetic order exists at these high temperatures as supported by the spin-lattice coupling. From these encouraging results we suggest to perform similar experiments on systems, where spin – spin interactions are mediated by bridging oxygen ions which through their anomalous polarizability induce such a coupling.

Below 100K another anomaly becomes evident in the χT plot which is inconsistent with the Curie-Weiss law (Fig. 4). The origin of this pecularity is at present unknown and requires further detailed studies. We can, however, speculate that dynamically correlated ferromagnetic clusters form due to the increasing lattice coupling which supports ferromagnetic order.

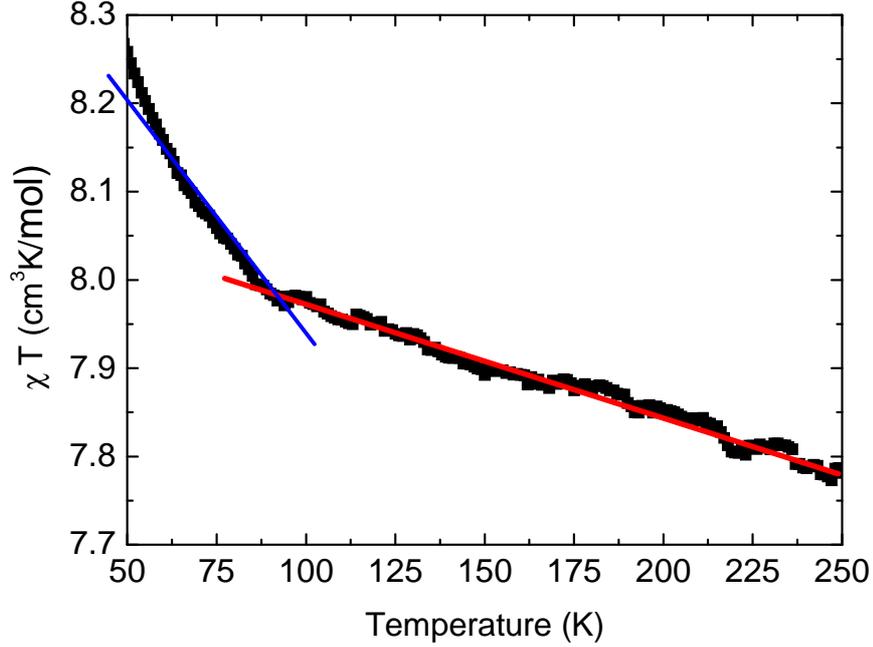

**Fig. 4** Temperature dependence of χT for the temperature range 50K < T < 250K. The blue line refers to the low temperature anomalous T-dependence, the red line is identical to the one in Fig. 3.

The lattice renormalization through the coupling term [Eq. (3)] appears dominantly in the soft optic long wave length mode. In the limit that $g_T$ is small, *i.e.* close to a polarizability catastrophe, this mode can be analytically expressed as:

$$\omega_{TO}^2(q=0) \cong \frac{1}{N}\left(\frac{2fg_T}{\mu} + \frac{2f}{m_2}\alpha\langle S_i S_j \rangle\right) \qquad (6)$$

where μ is the reduced cell mass, the spins are approximated by their thermal averages and $N = 2f + g_T + \alpha\langle S_i S_j \rangle$. While the first term resembles the bare soft mode frequency, the second term stabilizes this *at all temperatures* and inhibits a ferroelectric phase transition. Together with the fact that the extrapolated ferroelectric transition temperature is far in the negative temperature region, we suggest that quantum fluctuations do not hamper ferroelectric order, but it is hampered by the coupling to the spins. From Eq. (6) it is also obvious that upon the onset of antiferromagnetic order an additional increase in the soft mode frequency occurs which causes the drop of the dielectric constant at $T_N$ [3 – 5].

In summary, we have shown that the superexchange interaction between Eu 4f spins via the bridging oxygen ions induces a polarizability driven spin-lattice coupling which affects the spin and lattice dynamics in a subtle way. In particular, a temperature dependent enhancement of the magnetic susceptibility takes place which follows the soft acoustic mode temperature dependence. Our analysis together with the experimental data are to our knowledge the first to demonstrate that a structural instability can be detected by magnetic susceptibility measurements in a temperature regime far away from any long range magnetic order. This prediction together with the experimental observation offers a novel tool to analyze complex oxides with magnetic and structural transitions to detect possible spin-lattice couplings. Furthermore, the spin-lattice coupling supports short range ferromagnetic spin-spin correlations. In addition, the results show that ETO is not ferroelectric due to a suppression of the polar order by quantum fluctuations, but that the coupling to the Eu spins stabilizes the structure and inhibits long range polar order. These results question whether ETO can ever be a true multiferroic material in spite of the above demonstrated very strong spin-lattice coupling.

**Acknowledgement** ABH gratefully acknowledges helpful suggestions from K. A. Müller and discussions with A. Simon. This work is partly supported by the Swiss National Science Foundation.